\documentclass[aip, reprint, amsmath, amssymb, graphicx]{revtex4-1}

\draft 
\usepackage[utf8]{inputenc}
\usepackage{graphicx}
\usepackage{siunitx}
\usepackage{amsmath}
\usepackage{caption}
\usepackage{subcaption}
\usepackage{xcolor}
\usepackage[english]{babel}
\usepackage{csquotes}
\usepackage[utf8]{inputenc}
\usepackage[T1]{fontenc}
\usepackage{mathptmx}

\begin{document}
\title[Multifunctional operation of the double-layer ferromagnetic structure coupled by a rectangular nanoresonator]{
Multifunctional operation of the double-layer ferromagnetic structure coupled by a rectangular nanoresonator}

\author{Pierre Roberjot}
\affiliation{Institute of Spintronics and Quantum Information, Faculty of Physics, Adam Mickiewicz University, Pozna\'{n}, Uniwersytetu Pozna\'{n}skiego 2, 61-614 Pozna\'{n}, Poland}
\author{Krzysztof Szulc}
\email[Author to whom correspondence should be addressed: Krzysztof Szulc, ]{krzysztof.szulc@amu.edu.pl}
\affiliation{Institute of Spintronics and Quantum Information, Faculty of Physics, Adam Mickiewicz University, Pozna\'{n}, Uniwersytetu Pozna\'{n}skiego 2, 61-614 Pozna\'{n}, Poland}
\author{Jarosław W. K{\l}os}
\affiliation{Institute of Spintronics and Quantum Information, Faculty of Physics, Adam Mickiewicz University, Pozna\'{n}, Uniwersytetu Pozna\'{n}skiego 2, 61-614 Pozna\'{n}, Poland}
\author{Maciej Krawczyk}
\email[Author to whom correspondence should be addressed: Maciej Krawczyk, ]{krawczyk@amu.edu.pl}
\affiliation{Institute of Spintronics and Quantum Information, Faculty of Physics, Adam Mickiewicz University, Pozna\'{n}, Uniwersytetu Pozna\'{n}skiego 2, 61-614 Pozna\'{n}, Poland}

\date{\today}

\begin{abstract}
    The use of spin waves as a signal carrier requires developing the functional elements allowing for multiplexing and demultiplexing information coded at different wavelengths. For this purpose, we propose a system of thin ferromagnetic layers dynamically coupled by a rectangular ferromagnetic resonator. We show that a single and double, clockwise and counter-clockwise, circulating modes of the resonator offer a wide possibility of control of propagating waves. Particularly, at frequency related to the double-clockwise circulating spin-wave mode of the resonator, the spin wave excited in one layer is transferred to the second one where it propagates in the backward direction. Interestingly, the wave excited in the second layer propagates in the forward direction only in that layer. This demonstrates add-drop filtering, as well as circulator functionality. Thus, the proposed system can become an important part of future magnonic technology for signal routing.
\end{abstract}

\maketitle


A modern society is experiencing a rapidly-growing demand for interconnected wireless facilities, known as Internet of Things (IoTs),\cite{Mattern2010,Lin2017} which requires the development of faster ways to communicate and lower energy consumption to satisfy the environmental sustainability.\cite{Roselli2015} IoTs are devices receiving and sending back information using microwaves, thus the spin waves (SWs) are perfect candidates to be used in these developments.\cite{Mahmoud2020} It is due to the same range of frequencies as microwaves and a few orders shorter wavelengths, enabling miniaturization down to nanoscale. The SWs allow processing both the analog and digital signals at low energy cost, being inductively coupled to microwaves and compatible with CMOS technology.\cite{chumak2015,Kruglyak2010,Demokritov2013}


A possibility of frequency-dependent SW routing in a multiport device is essential to build any complex magnonic system.\cite{Khitun_2010} Therefore, implementation of the magnonic counterpart of the channel add-drop filter, which was demonstrated in photonics,\cite{Fan1998,Manolatou1999} is highly desirable. The crucial element of the add-drop filter is the resonator, e.g., a dot, a stripe or a ring, where the resonant modes mediate the coupling between the waves in unconstrained conduits. The resonance related to the fundamental mode of the stripe is widely explored to control the SW propagation and the interesting effects related to the chirality of the magnetostatic-stray-field coupling have been already demonstrated.\cite{Au_2012,Yu_2019,Chen_2019} From the other side, the ring-shaped resonators, extensively investigated in photonics, have been tested as a coupler between two SW waveguides, only recently.\cite{Wang2020} However, a single solid element with the circulating modes for SW routing remains unexplored. 

In our paper, we investigate theoretically a system composed of two Co ferromagnetic films coupled through a multimode resonant element---a stripe of the rectangular cross-section made from Py (Ni$_{80}$Fe$_{20}$). We observe various kinds of SW eigenmodes in the stripe. In particular, we found resonance modes in the form of single and double, clockwise (CW) and counter-clockwise (CCW), circulating waves. Our study goes along the exploitation of these modes for various functionalities, including add-drop filtering and a circulator. Our study is in line with recent activities in magnonics aiming in prototyping functional magnonic devices, like diodes,\cite{Grassi2020} circulators,\cite{Szulc2020} couplers,\cite{Wang2018,Wang2020-2} multiplexers,\cite{Heussner2020} and logic gates.\cite{Klingler2014,Zografos2017,Rana2018,Ustinov2020,Mahmoud20202} 

The paper is organized as follows. First, we present the geometry of the structure. Then, we describe the dispersion relations of the subsystems and analyze the eigenmode spectra. Finally, we demonstrate the operation of the system at different frequencies and present a summary.



\begin{figure}[!t]
    \includegraphics{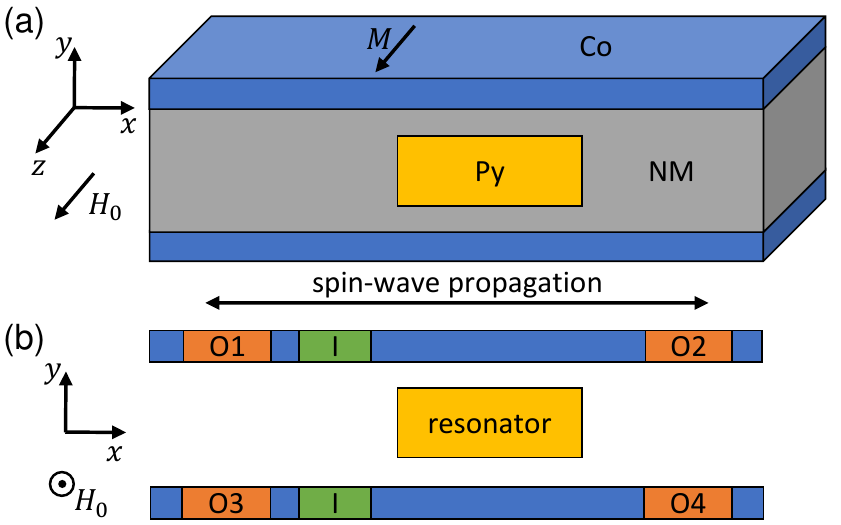}
    \caption{(a) The geometry of the bilayer structure under investigation. The system consists of two Co layers and a Py stripe separated by a nonmagnetic material. The separation between the layers and the stripe is $s = 10$ nm, the thickness and the width of the stripe is 50 nm and 100 nm, respectively, and the thickness of the layers is 5 nm. (b) Schematic representation of the four-terminal device. The possible positions for the inputs, i.e., the sources of SWs, are marked in green, while the four outputs in gray.}\label{fig1}
\end{figure}

We consider a system composed of two parallel, infinitely extended in the $xy$-plane, 5-nm-thick Co films and a resonant element in the form of a Py stripe of the thickness of 50 nm and the width of 100 nm placed between the films and oriented along the $z$-axis, as presented in Fig.~\ref{fig1}(a). The stripe is separated from the films by the 10-nm-thick nonmagnetic spacers.
The system is magnetized by a uniform in-plane bias magnetic field of magnitude $\mu_0H_0 = 0.05$ T directed along the stripe and perpendicularly to the direction of SW propagation. We assume the magnetization saturation and the exchange constant of Co to be $M_S = 1100$ kA/m  and $A = 20$ pJ/m, respectively, and for Py $M_S = 850$ kA/m and $A = 13$ pJ/m. The gyromagnetic ratio for the whole structure is $\gamma = 175.95$ rad/(sT). We consider the Damon-Eshbach geometry, where the SW group velocity and the magnetostatic coupling are significant.\cite{chumak2015} 

We consider the system shown in Fig.~\ref{fig1}(a) as a four-terminal magnonic device, from which one is selected as an input (I1 or I3) and the others as outputs (O1, O2, O3, O4), as shown in Fig.~\ref{fig1}(b). 
We want to guide the signal between pair of selected terminals through the stripe element by proper tuning the frequency of the excited SW. 

To describe the propagation in the system, we solve numerically the Landau-Lifshitz-Gilbert equation 
\begin{equation}
    \frac{\partial\textbf{M}}{\partial t} =-\gamma \mu_0 \textbf{M} \times\textbf{H}_{\text{eff}} + \frac{\alpha}{M_S}\textbf{M} \times \frac{\partial\textbf{M}}{\partial t}, \label{Eq:LLG}
\end{equation}
where $\textbf{M} = (m_x,m_y,m_z)$ is the magnetization vector, $\mu_0$ is the magnetic permeability of vacuum, $\alpha$ is the dimensionless damping parameter, and $\textbf{H}_{\text{eff}}$ is the effective magnetic field:
\begin{equation}
    \textbf{H}_{\text{eff}} =\textbf{H}_0 + \textbf{H}_{\text{m}} +\textbf{H}_{\text{ex}},
\end{equation}
which is a sum of the external magnetic field \(\textbf{H}_0= H_0 \hat{z}\), the magnetostatic field $\textbf{H}_\text{m}= -\nabla \varphi$, and the exchange field $\textbf{H}_{\text{ex}}= \frac{2A}{\mu_0M_S^2}\nabla^2\textbf{M}$. 
$\varphi$ is the scalar magnetic potential that fulfill Maxwell equations in the magnetostatic approximation:
\begin{equation}
    \nabla^2 \varphi=\nabla \cdot \textbf{M}.
\end{equation}
We linearized Eq.~(\ref{Eq:LLG}), assuming the harmonic time dependence $\exp(-i\omega t)$ and $H_0$ saturating the sample. We split the magnetization and the magnetostatic field into the static components parallel to the $z$-axis, $m_z=M_S$, $H_{\text{m},z}=0$, and the dynamic components laying in the $xy$-plane, $\textbf{m}=[m_x,m_y,0]$, $\textbf{h}_\text{m}=[-\partial_x\varphi,-\partial_y\varphi,0]$.
With these approximations, we have performed the finite-element method (FEM) simulations of the magnetization dynamics using COMSOL Multiphysics.

First, we solve eigenproblem based on linearized Eq.~(\ref{Eq:LLG}) using the frequency-domain solver in order to obtain the dispersion relation for SWs. In these simulations, the Floquet-Bloch boundary conditions are applied along the $x$-direction at the edges of the unit cell, which reproduce the effect of infinite layers. We neglect damping in these studies.

In the second approach, we solve linearized Eq.~(\ref{Eq:LLG}) in the time domain and for the finite structure. We used this approach to demonstrate functionality of the proposed devices. We excite the SWs in the input port with an antenna positioned 20 nm from the stripe edge. The antenna generates a dynamic magnetic sinusoidal signal of small amplitude
\begin{equation}
    S_{\text{ant}}(t)= 10^{-7}\gamma \mu_0 M_S^2 \sin{(2\pi f_0 \,t)}
\end{equation}
where $f_0$ is the frequency of excitation and the antenna's width $w_{\text{ant}} = \pi/k_{z0}$, with $k_{z0}$ being a wavenumber at the excitation frequency determined numerically from the dispersion relation obtained in frequency-domain simulations. In order to avoid reflection of the SWs from the system edges, we assume a linearly increasing damping at the edges of the Co layers.
The excitation last 3 ns, which is sufficient to observe the SW propagation and a resonant behavior, if present.


\begin{figure}[!t]
    \includegraphics{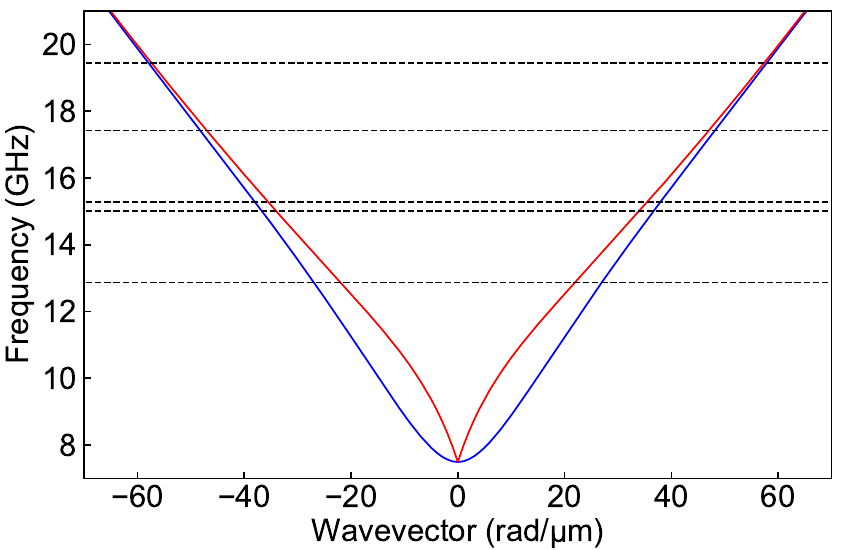}
    \caption{Dispersion relation of SWs in the subsystem composed of two Co layers separated by 70-nm-thick nonmagnetic material (red and blue solid lines) and the eigenfrequencies of the the Py stripe of 100-nm width and 50-nm thickness (horizontal dashed black lines).} \label{Fig:2}
\end{figure}



In order to understand the impact of the stripe on the propagation of the SWs in bilayered structure, we calculate the dispersion relations and the eigenmode spectra for two additional systems. The first one is made of two 5-nm-thick Co layers, separated by 70-nm-thick nonmagnetic material. The second one is a single Py stripe of width 100 nm and thickness 50 nm, i.e., the system in Fig.~\ref{fig1}(a) without the Co layers. 

Dispersion relation of the Co bilayer without a resonator is shown in Fig.~\ref{Fig:2}. There are two bands with the frequency difference proportional to the dynamical coupling between the SWs in the layers.\cite{Grunberg1981, Graczyk_2018,Szulc2020} The low- (blue) and high-frequency (red) modes can be assigned to antisymmetric and symmetric oscillations in the top and the bottom Co layer, respectively. 
We notice that these branches are getting closer with the increase of the frequency (and the wavevector), which means the coupling between the two layers is getting weaker with decreasing wavelength of the propagating waves.
This magnetostatic coupling between layers means that the SW excited in one layer will transfer between the layers periodically during the propagation. The distance on which the SW migrates between the layers is defined by the coupling strength. This effect was exploited in the directional couplers.\cite{Sadovnikov2015,Wang2018,Wang2020-2,Graczyk_2018,Sadovnikov2018}

\begin{figure}[!t]
    \includegraphics{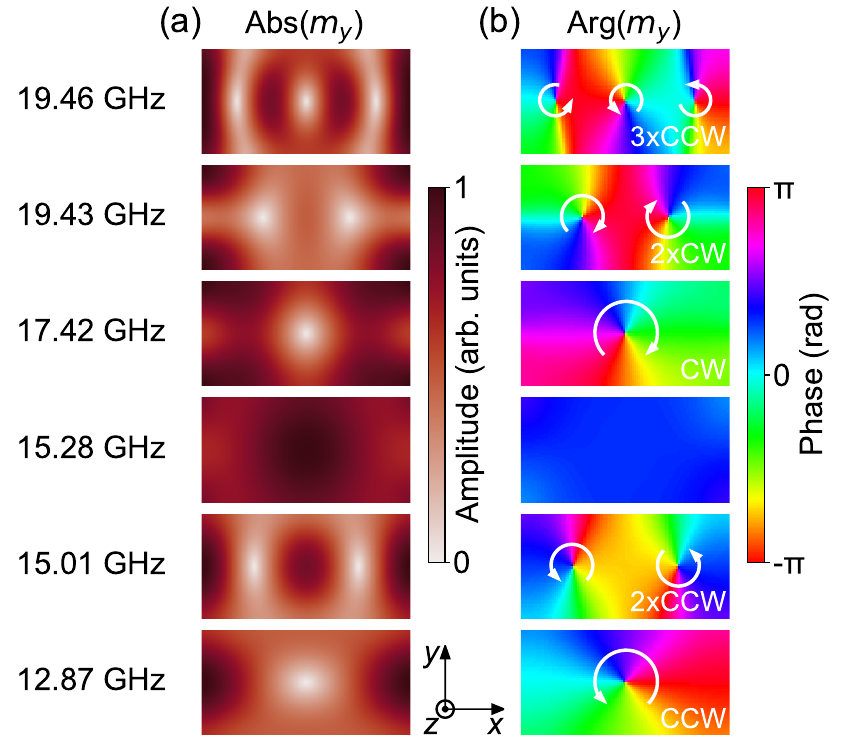}
    \caption{Amplitude (a) and phase (b) of the $m_y$ component of the magnetization in the isolated Py stripe for the modes shown by dashed black lines in Fig.~\ref{Fig:2}. The CW and CCW modes are indicated in (b), the continuous phase change and two zeros of the amplitude at 15.01 and 19.43 GHz indicate a double CCW and CW circulating modes, respectively.}
    \label{fig3}
\end{figure}

The spectrum of the eigenmodes of the second subsystem is presented in Fig.~\ref{Fig:2} with dashed black lines and indicates SW resonant modes of the infinitely-long Py stripe with the magnetization saturated along the $z$-axis. In the investigated frequency range, the six modes are present with two pairs of the modes very close in frequency, at about 15 and 19 GHz.
To analyze the types of resonant excitations, we plot in Fig.~\ref{fig3} the amplitude and phase of the $y$-component of the magnetization, Abs($m_y$) and Arg($m_y$), respectively. We can see a fundamental mode with the in-phase oscillations in the stripe cross-section at 15.28 GHz. 
Interestingly, all other modes are CW or CCW circulating SWs, as indicated by continuous change of the SW phase in Fig.~\ref{fig3}(b), and marked with the white arrows. We found also the modes composed of the two CCW (at 15.01 GHz) or two CW (19.43 GHz) circulated oscillations, we will call them double-CCW and double-CW circulating modes, respectively. All considered CCW modes have lower frequency, than the corresponding CW modes, which points at the presence of a nonreciprocity in this system. As shown in Fig.~\ref{fig3}(b), the double circulating modes present circulations possessing relative phase shift by 180$^\circ$ at the top and the bottom edges of the stripe. 
These two properties shall cause different coupling of the circulating resonant modes to the modes propagating in the same direction in the upper and lower ferromagnetic layers or alternatively, a different coupling between the modes propagating in the same layer, but in the opposite directions. Indeed, this feature is used to control waves in the waveguides coupled with the ring resonators and whispering gallery modes in photonics\cite{Rowland_1993,Little_1997,Bogaerts_2012} and magnonics\cite{Odintsov_2019,Wang2020}. We will use this property for demonstration of the SW circulator.

The gradient of the phase on the sides of the resonator depends on the number of the circulation areas, it is one for single CW/CCW and two for double-CW/double-CCW modes [see, Fig.~\ref{fig3}(b)]. We expect that if the gradient of the phase at the top and bottom edge of the resonator match to the wavenumber and the phase change of the SW propagating in the ferromagnetic layer, there will be an enhanced transmission of SW in one and suppressed in the opposite direction of propagation.

Before studying the transfer of SWs between Co layers at frequencies close to the circulating resonances of the Py stripe, we analyzed also the dispersion relation of the base system, i.e., bilayered structure with the Py stripe in-between, with periodic boundary conditions along the $x$-axis. The spectra (shown in the Supplementary Material) has many band gaps between more or less dispersive bands and it is significantly different from the spectra of the two Co layers shown in Fig.~\ref{Fig:2}. It indicates a strong and complex coupling between the propagating modes in the bilayer and the resonant modes of the Py stripe. 

\begin{figure*}[!t]
    \includegraphics[width=0.8\linewidth]{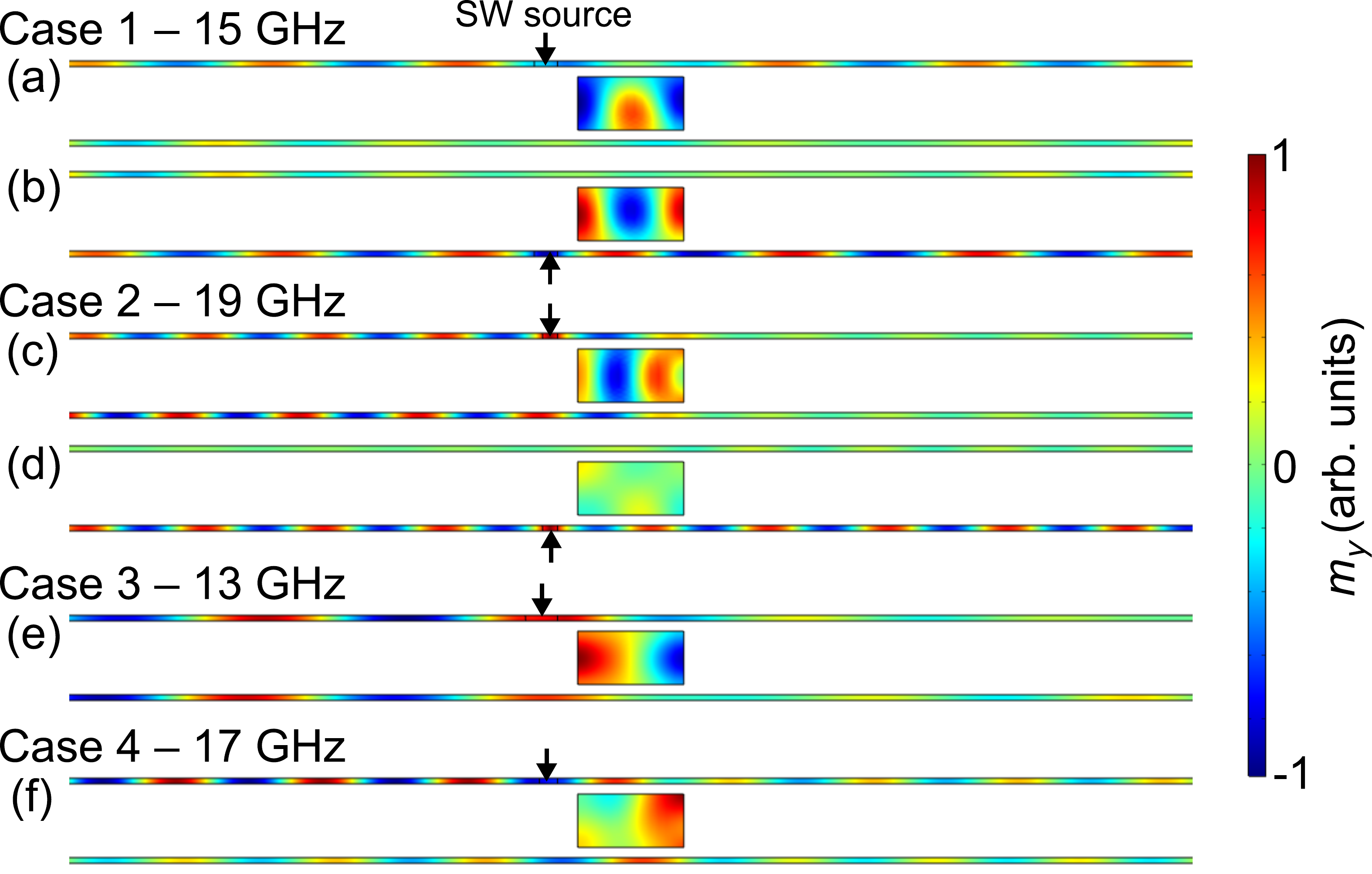}
    \caption{Spin-wave amplitude plots showing propagation in the investigated structure at selected excitation frequencies to demonstrate different functionalities. The wave excited in the top (a) or bottom (b) layer at 15 GHz weakly couples with the resonator. The observed transfer of the wave from the top to the bottom layer is an effect of the direct coupling between the Co layers. This is case 1. (c) The excitation at 19 GHz in the top-left part of the layer is transferred via resonator to the bottom layer and propagates to the left. (d) The excitation is located in the bottom layer on the left, the wave propagates straight to the right without coupling. This is operation of the magnonic add-drop filter and circulator, case 2. (e) The wave excited at 13 GHz is transferred through the resonator to the bottom layer and propagates in both direction, case 3. (f) The wave excited in the top layer at 17 GHz is reflected back from the resonator, case 4. The simulations were performed in the time domain, the plots show the SW amplitude, i.e., $m_y$ component of the magnetization, after 2 ns from excitation.}
    \label{fig4}
\end{figure*}

To demonstrate how different resonances in the Py stripe influence the coupling and transfer of SWs between the Co layers, we perform time-domain simulations for the four cases. Case~1 at 15 GHz, it is around the resonance of the fundamental and the double-CCW mode. For the case~2, we selected the frequency 19 GHz, which is close to the double-CW circular mode of the stripe (mode at 19.43 GHz in Fig.~\ref{fig3}). Case~3, 13 GHz, was chosen to be close to the single-CCW circulating mode of the stripe (12.87 GHz in Fig.~\ref{fig3}), and the case~4 at 17 GHz, the frequency close to the single-CW circulating mode (mode 17.42 GHz).

\textbf{Case 1} at 15 GHz is shown in Fig.~\ref{fig4}(a) and (b) for the SW source placed in the top and the bottom layer, respectively. The SW is transferred to the Py resonator but then transfers back to the Co layer, propagating in the same direction. We can observe two phenomena. First, the SW changes its phase upon the movement through the resonator. Second, the transfer between Co layers is delayed. 
During propagation, the SW transfers between the layers coupled by the stray dynamic magnetic field, similarly to the directional couplers.\cite{Sadovnikov2015,Sadovnikov2017,Wang2018,Graczyk_2018} Such behavior is present also at low frequencies, i.e., below the fundamental mode of the resonator, where the dynamical magnetostatic coupling between Co layers is strongest and thus the period of the transfer between the layers is shortest.

\textbf{Case 2, add-drop filtering and circulator.} In this case, we found a functionality analogous to the circulator, as demonstrated in Fig.~\ref{fig4}(c) and (d). In this device, the wave excited in one terminal has to be transferred only to one of remaining terminals according to the assumed clockwise or counter-clockwise circulation rule.\cite{Szulc2020} We found the 19-GHz frequency optimal for such a functionality, close to the double-CW circulating mode resonance of the Py stripe (Fig.~\ref{fig3}, mode at 19.43 GHz). At this frequency, the wave propagating from the antenna can excite the resonant mode of the stripe which will act as an energy harvester due to the magnetization circulating mode. Indeed, when the wave is excited on the left side of the top layer (the input port I1) [Fig.~\ref{fig4}(c)], the Py resonator redirects it to the bottom layer as a wave propagating to the left, to the output port O3. Importantly, an isolation of the ports O2 and O4, exceeding 5 times of the signal directed to O3, exists over 600 MHz range.
When the input port is placed in the bottom layer on the left (I3) from the resonator [Fig.~\ref{fig4}(d)], the wave propagates directly to the port O4 in the same layer. For the wave excited in I4, the resonator redirect to O2, while excited in I2 will propagate to O1. We observe similar functionality also at 13 GHz but with larger leakage of the energy to the other ports. The presented operation can be used for a design of magnonic circulator or add-drop filtering.  

\textbf{Case 3} at 13 GHz is shown in Fig.~\ref{fig4}(e). The wave introduced in the port I1 couples with the single-CCW circulating mode in the Py stripe which allows to make transfer to the bottom layer and form waves propagating to the both ports located there, O3 and O4. Although the intensity of the waves reaching the ports O3 and O4 differs, the clear splitting of the wave propagating in the top layer has been demonstrated. 

\textbf{Case 4.} In this case, we introduce the wave at 17 GHz in the port I1. As can be seen from the Fig.~\ref{fig4}(f), the excited wave is transferred to the Py stripe and the circulating mode passes it to different ports but mostly go back to the top-left output O1. The Py resonance involved in this operation is related to a single-CW circulating mode, thus the functionality can be expected to be similar to the coupling with the ring resonators \cite{Wang2020} or of the circulator shown in the case 2. However, the mismatch between the wavelength of the propagating SW and the resonance length results in a weak coupling and unclear operation. 

Interestingly, the coupling of SWs excited in the left part of the bottom Co layer (I3 port) with the Py stripe is suppressed at all considered frequencies. We attribute this to the unidirectional magnetization precession and the chirality of the magnetostatic stray field coupling\cite{Au_2012,Yu_2019}.


The investigations presented above were made without damping in order to have clear visualization of the coupling processes. We performed also the same simulations but with a damping, we assume in Co layers $\alpha_{\text{Co}} = 0.01$ and in Py stripe $\alpha_{\text{Py}} = 0.005$. We observed the presence of the same phenomena only with an attenuation. The results are presented in the Supplementary Material.

Using a stripe of 100-nm width and 50-nm thickness, we have demonstrated some of the possibilities to control the propagation of the SWs in thin ferromagnetic films at different frequencies. The energy plots shown in the Supplementary Material additionally support these demonstrations. However, the demonstrations were done for selected materials, the one relative magnetization orientation and at the fixed geometrical parameters. The functionality of the resonance element depends on the wavelength of the propagating wave and its relation to the size of the stripe, and chirality of the magnetization precession, thus further investigations and model developing is necessary to optimize the operation of the proposed solutions. 


In conclusion, we explored numerically the SW behavior in a structure composed of two 5-nm-thick Co layers separated by a rectangular Py stripe, all homogeneously magnetized parallel to the stripe axis. We show that its dispersion relation is significantly different as compared to a bilayer structure due to dynamical coupling of propagating SWs in Co layers with the resonant modes of the Py stripe. Interestingly, the low-frequency part of the isolated-stripe spectra consists of the fundamental mode and a number of single and double, non-degenerated CW and CCW circulating modes. The interaction between propagating SWs in layers and modes circulating in the resonator allows for design add-drop filters and circulators for SWs. 

In particular, we show that depending on the direction of the SW propagation and on the excitation frequency, the SWs can transmit in four possible routs: direct propagation between the two Co layers, circulation (SWs are propagating from one layer to the other keeping the direction of propagation), reflection (forth and back to the excitation point), and uncoupled propagation (SWs are propagating in the one layer only). Especially effective is the use of double-CW circulating mode of the stripe allowing us to demonstrate the functionality of the magnonic circulator. This shows that the circulating modes in the ferromagnetic stripe are promising for design of signal-processing magnonic devices---especially multiplexers and demultiplexers.

\begin{acknowledgments}
The research leading to these results has received funding from the Polish National Science Centre, project no. UMO-2018/30/Q/ST3/00416.
\end{acknowledgments}
 
 The data that support the findings of this study are available from the corresponding author upon reasonable request.
 
\bibliography{literature}

\end{document}